\documentclass[lettersize,journal]{IEEEtran}
\usepackage{amsmath,amsfonts}
\usepackage{algorithmic}
\usepackage{algorithm}
\usepackage{array}
\usepackage[caption=false,font=normalsize,labelfont=sf,textfont=sf]{subfig}
\usepackage{textcomp}
\usepackage{stfloats}
\usepackage{url}
\usepackage{verbatim}
\usepackage{graphicx}
\usepackage{cite}
\usepackage{caption}
\usepackage[numbers,sort&compress]{natbib}
\begin{document}
\title{Space-Air-Ground Integrated Networks for\\
6G Mobile Communications}


\author{Tianming Lan
\thanks{The author is with Nanyang Technological University, Singapore. Email: tianming001@e.ntu.edu.sg.}
}



\maketitle

\begin{abstract}
After the industrialization of 5G cellular communications, 6G has increasingly become a research hotspot in the academia. Space-Air-Ground Integrated Network (SAGIN) is a key supporting technology for 6G because of its advantages such as high-speed transmission and expanded coverage. This paper summarizes the motivation to develop the SAGIN-assisted 6G first and introduces the current situation of SAGIN. We then try to discuss the design concept of the SAGIN-assisted 6G, and list the problem and challenges. Moreover, we propose an architecture of the SAGIN-assisted 6G and identify a series of key technologies that are needed in different layers. Finally, in order for readers to better understand our hierarchical architecture, we use a case study discussing the congestion problem in the Integrated layer.
\end{abstract}


\section{Introduction}
Mobile communication has developed from the first generation of mobile communication (referred to as ``1G") in the 1980s to the fifth generation of mobile communication (referred to as ``5G"). The terrestrial mobile communication represented by 5G can provide diverse service support capabilities and a good user experience, but there are still some problems existing. In addition, traditionally, ground mobile communication networks and satellite communication networks have been independently developed, constructed and deployed, and cannot provide air-space-ground integrated resource control and business continuity service capabilities.\par
In the 6th generation of mobile communication (referred to as ``6G"), the coverage of mobile communication networks is expected to be greatly expanded, not limited to about 20\% of the land area and 6\% of the earth's surface area covered by terrestrial mobile communication, and will also involve the ocean, mountains, forests, deserts and other wider areas that cannot provide signal coverage through terrestrial cellular base stations. Therefore, terrestrial mobile communication and satellite communication should complement each other. It is the general trend for the two to jointly build a satellite-ground fusion network covering the whole world. Among them, inland densely populated areas are covered by terrestrial mobile communication networks, giving full play to the advantages of capacity and meeting the demand for massive connections; while remote areas and oceans that cannot be covered by terrestrial mobile communication are covered by satellite communication, which can give full play to the coverage advantages of satellite communication, saving ground base station construction and operation and maintenance costs. It can be expected that 6G mobile communications will build a three-dimensional coverage Space-Air-Ground Integrated Network (SAGIN) based on ground cellular communications, satellite networks and high-altitude platforms.\par
The present paper provides a comprehensive analysis of the SAGIN and 6G from multiple perspectives. Previous research efforts \cite{lin2021path, mishra2021drone, xiao2022leo, zhang20196g} have also examined the SAGIN and 6G; however, their analyses have been found to be inadequate. Specifically, Lin~\emph{et~al.}~\cite{lin2021path} focused on the technology involved in service level and standardization, whereas Mishra~\emph{et~al.}~\cite{mishra2021drone} investigated the role of drones in the context of SAGIN. Xiao~\emph{et~al.}~\cite{xiao2022leo} solely discussed the challenges and issues associated with SAGIN, and Zhang~\emph{et~al.}~\cite{zhang20196g} explored the structural and interface design of SAGIN. By contrast, this paper offers an examination of the underlying motivations for developing the SAGIN-assisted 6G and proposes a novel design concept for the SAGIN-assisted 6G. Furthermore, the key technologies employed across all layers of the OSI network model are summarized in this paper.\par
The structure of this paper is as follows. We firstly introduce the motivations for researching and developing the SAGIN-assisted 6G in Section \ref{motivation}. Then Section \ref{overview} summarizes the current development status of SAGIN. In Section \ref{desn_concept}, we present a set of design principles for the SAGIN-assisted 6G. Then in Section \ref{pro_cha}, we summarize the problems and challenges that may be encountered in the development of the SAGIN-assisted 6G. As a step further, Section \ref{key_tech} introduces the key technologies that the SAGIN-assisted 6G may use based on section \ref{desn_concept}. In order for readers to better understand the above key technologies, Section \ref{case} uses an example to illustrate the routing technology in them. Finally we conclude the paper in Section \ref{conc}.\par

\section{Motivation}\label{motivation}
Development of the SAGIN-assisted 6G is gaining traction within many companies and organizations, such as 3rd Generation Partnership Project (3GPP), International Telecommunication Union (ITU), and China Communications Standards Association (CCSA). Many of them proposed the demand and key technologies of SAGIN on their white paper.\par
Electronics and Telecommunications Research Institute (ETRI) published a 6G white paper that introduced the three-dimensional coverage demand of SAGIN. Japan NTT DOCOMO introduced the space coverage demand of SAGIN in their white paper \cite{JapanNTT}. 3GPP proposed the key integrated technology of SAGIN in TR38.81 Release 16. Above all, the main motivation for developing the SAGIN-assisted 6G comes from the following points.\par

\subsection{Coverage}
The envisioned objective for future global communication networks is to provide reliable and high-speed services to users from all locations, customized to their unique needs and available at any time. This is also the ultimate goal for the development of the 6G network.\par

Nonetheless, the current terrestrial mobile network infrastructure appears to fall short of meeting the demands of communication. With a population coverage rate of approximately 70\%, terrestrial mobile communication services cater to a mere 20\% of the total land area, which is less than 6\% of the entire surface area of the earth. This limitation renders seamless global coverage a formidable challenge. However, satellite communication boasts the advantages of extensive coverage, high transmission capacity, and terrain-independent operability, providing a compelling case for the increasing importance of SAGIN as a means of expanding the application scenarios of terrestrial mobile communication. Therefore, SAGIN is expected to play a crucial role in shaping the infrastructure of the upcoming 6G networks.\par

\subsection{Technology}
Traditionally, satellite communications and terrestrial communications use different technical systems. But with the rapid development of terrestrial mobile communication, there are more and more demands for communication between terrestrial networks and satellites. Different technical systems have greatly hindered the development of communication services between satellite and terrestrial networks. It's very necessary and emergent to develop a unified and complete technical system, which can support network construction and operation and maintenance in both satellite and terrestrial networks.

\subsection{Market}
Satellite Internet has huge market prospects, and many capitals have entered recently. In the Satellite Internet Market Analysis Report of Grand View Research \cite{grandviewresearch}, the global market for satellite internet, which was valued at USD 7.37 billion in 2021, is anticipated to increase at a CAGR of 11.6 percent between 2022 and 2030.\par
The Satellite Internet market is divided into energy and utility, public administration, transportation and cargo, maritime, defense, corporations/enterprises, media and broadcasting, and other categories based on industry. In 2021, this market was dominated by the government and public sector, with a share of 19.8\% \cite{grandviewresearch}. To link every area of their countries to the broadband network, governments all over the world have been investing in satellite broadband technologies.  They also rely on state-of-the-art technology to provide broadband services in rural areas.\par


Additionally, a unified industrial chain (including terminals, networks, test equipment, etc.) is also needed for network operators to provide a better user experience. Furthermore, the industry needs unified resource management and control, system operation and maintenance, unified business service management, and user experience management for both satellite communications and ground mobile communications to greatly reduce the construction and operation, and maintenance costs of satellite communications.

\section{An Overview of SAGIN}\label{overview}

\subsection{Satellite Constellations}
About 20 years ago, the idea of satellite constellation appeared and the exemplary pioneer includes Iridium \cite{fossa1998overview} and
Globalstar.
So far, the academic community has established an unequivocal description of a satellite constellation. It refers to an amalgamation of various satellites that share comparable characteristics and purposes, which are situated in congruous or complementary orbits, and collaborate harmoniously to accomplish designated missions under shared control.
The final version of Iridium is 66 satellites with 6 orbital and 11 satellites inside. For the Globalstar constellation, 48 satellites make up it, which is located at a height of around 1400 kilometers. They offered worldwide 2G mobile communication links. However, they only got limited success mainly due to service costs.\par
After their failure, satellite communications stagnated for a long time until the small satellites appeared.
The technology miniaturization has resulted in a significant reduction in service costs, thereby paving the way for new opportunities. As a result, there has been renewed interest in utilizing satellite constellations as a means of providing connectivity from space in light of the emerging market opportunities.
As more and more satellites are launched into space, a number of companies have proposed satellite constellations with thousands of Low Earth Orbit (LEO) satellites. And such a satellite constellation composed of a large number of satellites is called ``mega-constellations". \par
The first to join the field of mega-constellations is Iridium NEXT, which started launching satellites in 2017 and has now formed an LEO network consisting of 75 satellites, providing ground internet service. Iridium is a representative polar-orbiting constellation. Each of its orbital planes has the same inclination and the same number of satellites which are evenly distributed. This type of constellation is characterized by dense satellites in high latitudes and sparse in low latitudes.\par
\begin{figure}[htbp]
    \centering
    \begin{minipage}{\linewidth}
        \centerline{\includegraphics[width=5.0cm]{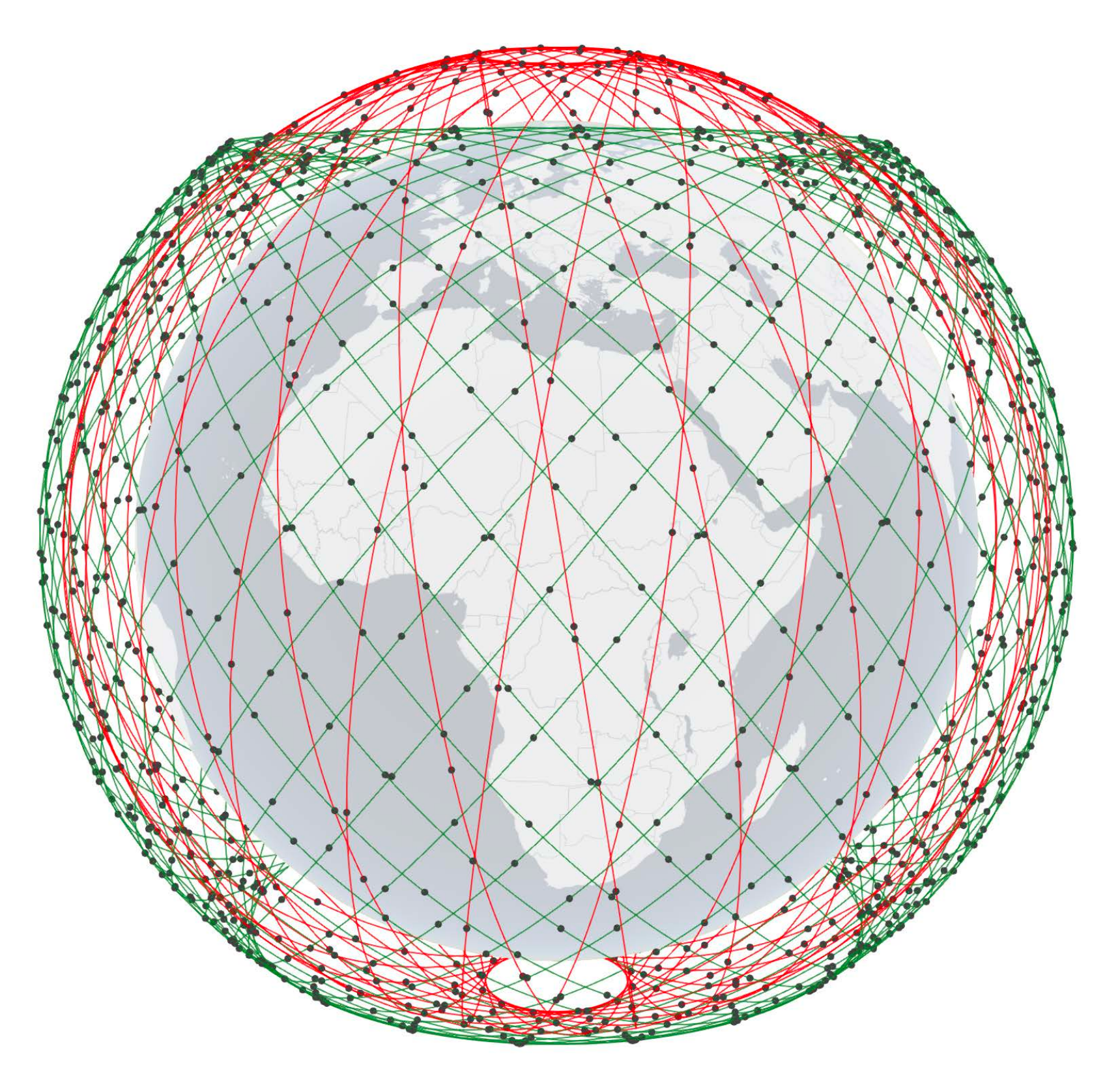}}
        \centerline{(a) Telesat constellation.}
    \end{minipage}
    \vfill
    \begin{minipage}{\linewidth}
        \centerline{\includegraphics[width=5.0cm]{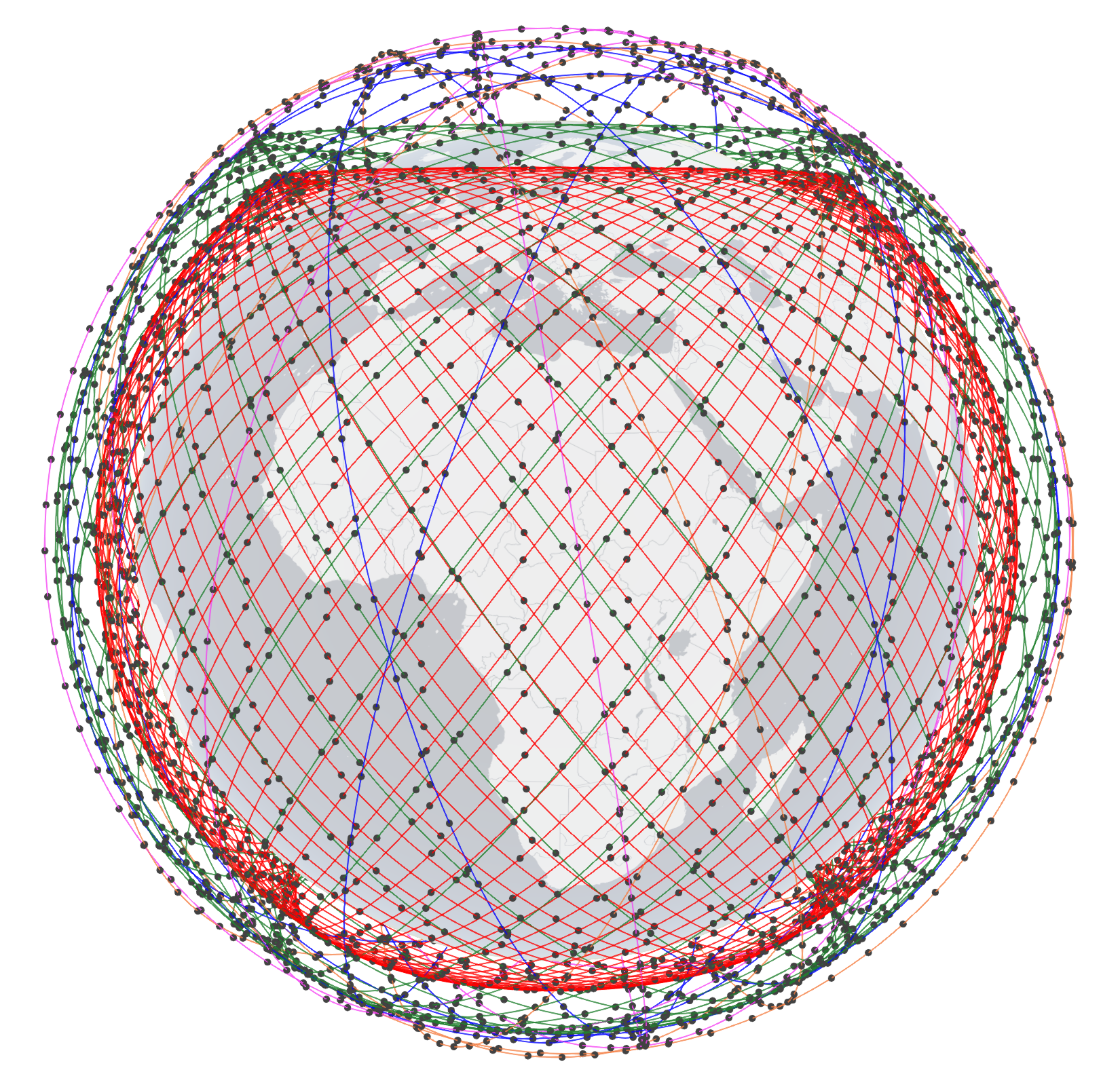}}
        \centerline{(b) Starlink constellation.}
    \end{minipage}
    \vfill
    \begin{minipage}{\linewidth}
        \centerline{\includegraphics[width=5.0cm]{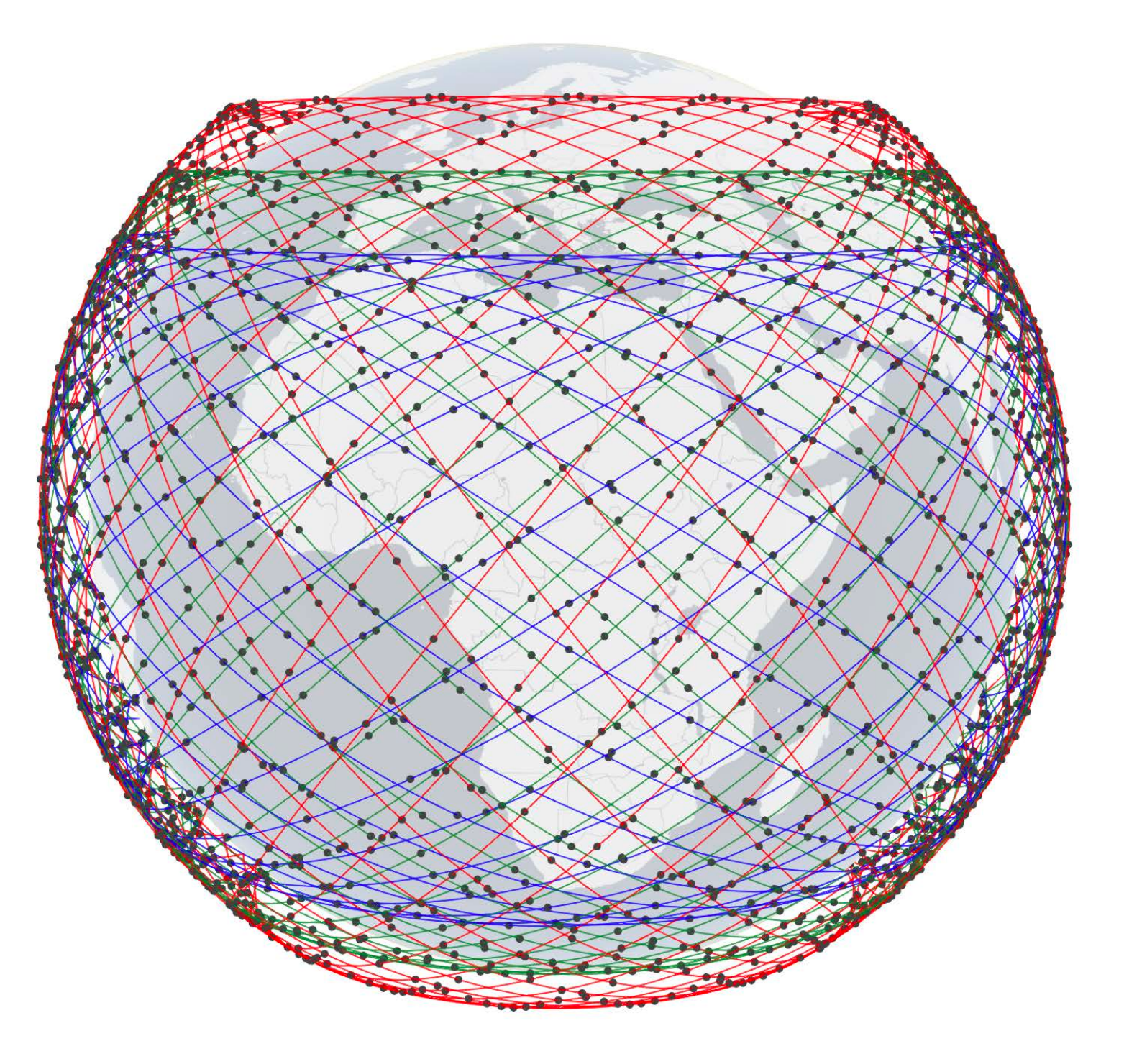}}
        \centerline{(c) Kuiper constellation.}
    \end{minipage}
    \vfill
    \caption{Example of Mega-constellation \cite{Leo_sat_net}.}
    \label{fig-conste}
\end{figure}

What joined the industry at the same time was Telesat. Telesat established a satellite constellation plan called Telesat Lightspeed \cite{lester1978telesat} and expanded this plan to 300 satellites in 2017, of which about 80 are polar-orbit satellites, and the rest are inclined orbit satellites. The schematic diagram of the topology is shown in Fig. \ref{fig-conste} (a).\par

This was followed by the very ambitious SpaceX's Starlink project \cite{mcdowell2020low}. SpaceX plans to launch 41,493 satellites in the future. The Starlink constellation does not use a polar orbit constellation like Iridium, but adopts a multi-layer combination constellation shown in Fig. \ref{fig-conste} (b). They are mainly based on inclined circular orbit constellations, supplemented by a small number of polar orbit satellites. The inclined circular orbit constellation is composed of circular orbits with the same height and inclination. But inclined circular orbit also has its disadvantages in that it can not cover high-latitude regions including the polar regions. Therefore, the Starlink constellation has added polar orbit satellites to make up for this shortcoming.\par
OneWeb entered the Satellite Internet market in 2019. They plan to build a polar circular orbit constellation of 720 satellites. The constellation consists of 18 orbital planes and each orbital plane has 40 evenly distributed satellites with an altitude of 1200 km.\par

The Kuiper \cite{10.1145/3419394.3423635} constellation of Amazon has attracted quite a lot of attention in recent years, because Amazon announced that they would put Amazon's cloud services on the Kuiper and also cooperate with the data centers that have been deployed around the world to improve cloud computing capabilities. The Kuiper project plans to launch 3,236 satellites in the future \cite{10.1145/3419394.3423635}. As shown in Fig. \ref{fig-conste} (c), these satellites are distributed in three inclined circular orbit constellations with different orbital layers.

\subsection{Satellite Capacity}
Initially, Iridium and Globalstar only provided communication as relay nodes, in which satellites didn't connect with each other. These satellites provided 2G mobile services and short data messaging to specific handset terminals, with a maximum data rate of 14.4 kbps.
But as technology developed, Inter-Satellite Links (ISLs) appeared, which greatly expanded the communication range of users. 
The ISLs of the Iridium NEXT use microwaves for communication and can provide bandwidth of Mbps or even Gbps.\par
Telesat uses lasers for the ISLs. The laser link can provide a bandwidth of more than 100Gbps. At the same time, because the propagation speed of the laser in a vacuum is faster than that in the optical fiber network on the ground, the upper bound of Telesat's global communication delay will be greatly reduced. They have established more than 50 ground stations around the world as relays of the satellite network, and the two satellites that cannot be directly connected rely on ground relays for data transmission. However, the laser link also has its disadvantages: it is difficult for two satellites to align and it takes a long time to establish. This makes satellite constellation topology design become difficult.\par 
Starlink also uses lasers as the propagation medium for the ISLs. And the difference of OneWeb is that it abandons the ISLs to pursue the stability of transmission, and the communication between satellites is completely completed by ground relay. The Kuiper constellation uses microwave as the communication medium between satellites in a different layer and uses the laser in the same layer.\par



\section{Design Principles for the SAGIN-assisted 6G}\label{desn_concept}
In response to the requirement of 6G SAGIN, 6G system design needs to satisfy the consistency between the terrestrial communication system and the space communication system. Besides, system design also needs to ensure that the system has different protocols for different scenarios.\par 
In our opinion, the SAGIN-assisted 6G system should have the following features.

\begin{itemize}
\item The system uses integrated air interface transmission waveform and multiple access multiplexing technologies which can be configured in different scenarios and it can provide stable data transmission, fast access to users.
\item The system has a unified frame structure and physical channel design which can be combined to optimization technology in special scenarios and support efficient transfers and different types of user services.
\item The system supports on-demand configuration of protocol tailoring and optimization, space-ground collaborative transmission, and seamless switching based on a unified SAGIN air interface protocol stack and network architecture.
\item This system adopts a unified spectrum allocation and management strategy to realize frequency resource sharing and interference avoidance facing various scenarios from space to ground.
\end{itemize}

The SAGIN-assisted 6G system also needs to support diverse services in various scenarios in Table \ref{tab:services}.

\begin{table*}[htbp]
\caption{The services support capability of the SAGIN-assisted 6G system.}
\centering
\begin{tabular}{|c|c|c|}
\hline
Services Type & Description\\
\hline
Mobile data/voice communication & Implement global seamless coverage\\
\hline
Relay communication & Provides high-speed direct connection to unreachable area\\
\hline
Backhaul communication & Provide high-speed backhaul for remote areas and aggregated IOT traffic\\
\hline
Switch management & Provide smooth and seamless switching for high-speed moving equipment\\
\hline
Broadcasting & Provide data, video, TV and other broadcasting services\\
\hline
Emergency communication & Emergency disaster early warning, emergency information reporting\\
\hline
\end{tabular}
\label{tab:services}
\end{table*}
The specific requirements of the industry for the SAGIN-assisted 6G are as follows: support larger cell coverage whose radius can reach more than 500 km; support the dynamic distribution of users whose height ranges from 10 to 20 km; support mobile users whose speed can be up to 1000km/h; the user experience rate is not less than 1 Mbit/s; the maximum spectral efficiency is greater than 3 bit/s/Hz.\par
\begin{figure*}[htbp]
  \centering
  \includegraphics[width=0.8\linewidth]{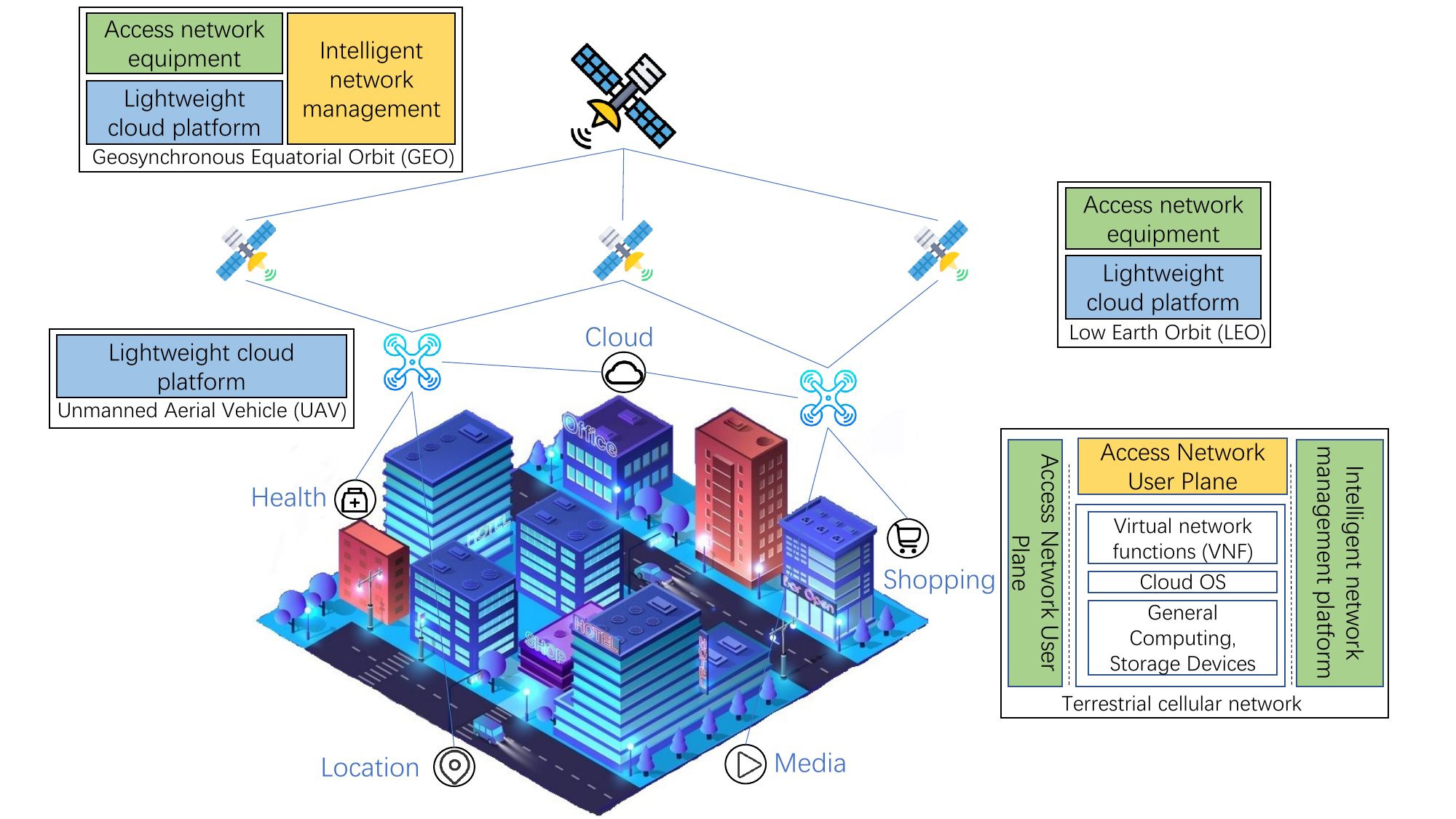}
  \captionsetup{justification=centering}
  \caption{An architecture of the SAGIN-assisted 6G system.}
  \label{fig:SAGIN_arc}
\end{figure*}

 
Above all, four principles that the SAGIN-assisted 6G should follow are given:
\begin{enumerate}
    \item The system should support global three-dimensional wide-area coverage, provide service continuity and QoS guarantee.
    \item The system should adopt a unified network architecture and interface protocol to support the deployment of all scenarios.
    \item The system should support the single-mode design of star-ground terminals based on the integrated design of air interface technology.
    \item The system should support technical feature tailoring and optimized design according to different scenarios and devices.
\end{enumerate}
Based on all the above descriptions, we give a design of the SAGIN-assisted 6G system in Fig, \ref{fig:SAGIN_arc}. This architecture can ensure collaborative management and control of SAGIN and flexible and on-demand deployment of network functions through the flexible division of network functions between satellites and ground.

\section{Problem and Challenge}\label{pro_cha}
As a brand new field, the SAGIN-assisted 6G will definitely face many difficulties and challenges. Here we summarize some representative problems.
\subsection{Space-ground link transmission problem}
\textbf{High latency, the effect of engaging in Doppler.} The communication distance of the space-ground link is in the range of hundreds to thousands of kilometers, which is far beyond the station distance of the terrestrial cellular mobile communication, which brings unavoidable high delay. At the same time, the fast movement of the access point will cause serious Doppler shift problems, and will also limit the use of MIMO multi-stream technology, affecting service continuity.\par 
\textbf{Space-ground coordination problem.} Space-ground coordinated coverage and transmission enhancement needs to be considered to solve space-ground and inter-satellite interference coordination and frequency sharing, which requires solving the problem of network node interface information interaction.
\subsection{Communication devices problem}
Generally speaking, communication equipment needs to adapt to the complex environment of land, sea, air and space, and needs to support multi-frequency and multi-mode. User terminals need to ensure portability, high energy efficiency, and be green and intensive. The biggest challenge is the reliability of satellite components. Heat dissipation capacity constraints require transmission waveforms with low peak to average power ratio (PAPR) characteristics, and such satellite platforms do not have sufficient data and signal processing capabilities. The above problems make the lightweight design of the protocol very urgent.\par 
\subsection{Air interface and network management problem}
A unified air interface technology system and resource sharing are important foundations for the realization of the SAGIN-assisted 6G.\par 
\textbf{Dynamic frequency sharing and interference cancellation.} The air interface links of the SAGIN-assisted 6G are more complex and diverse than the ground communication network nowadays. In SAGIN, every ISLs has a large data rate requirement, but the radio frequency resources are limited. Therefore, smart and efficient frequency sharing and cancellation methods are urgently needed.\par 
\textbf{Antennas and RF.} Current terrestrial mobile communication technologies cannot be directly used in the non-terrestrial networks. First, it needs to satisfy the space-to-ground coverage requirement. At the same time, compared with traditional communication satellites, the coverage method of LEO satellites in the era of mega-constellations has also changed, which adopts the method of spot beam multiplexing for coverage. Therefore, antenna beamforming and radio frequency technologies for non-terrestrial network access sites need to be redesigned before they can be used in space.\par 
\textbf{Intelligent access and mobility management.} The overall capacity of the satellite network is smaller than the terrestrial networks, and the capacity of a single satellite is limited. In order to ensure user experience, satellite communication systems need to design appropriate user access and handover strategies, which include selecting appropriate satellite beams and channels for users. At the same time, the movement speed of LEO satellites relative to the ground is about 27,000 km/h. Therefore, the service duration of each satellite is only tens of seconds, and one service may include multiple satellite and ground station switches. In addition, the switching problem also involves the switching between different communication systems.

\section{Key supporting technologies}\label{key_tech}
\begin{figure*}[htbp]
  \centering
  \includegraphics[width=0.8\linewidth]{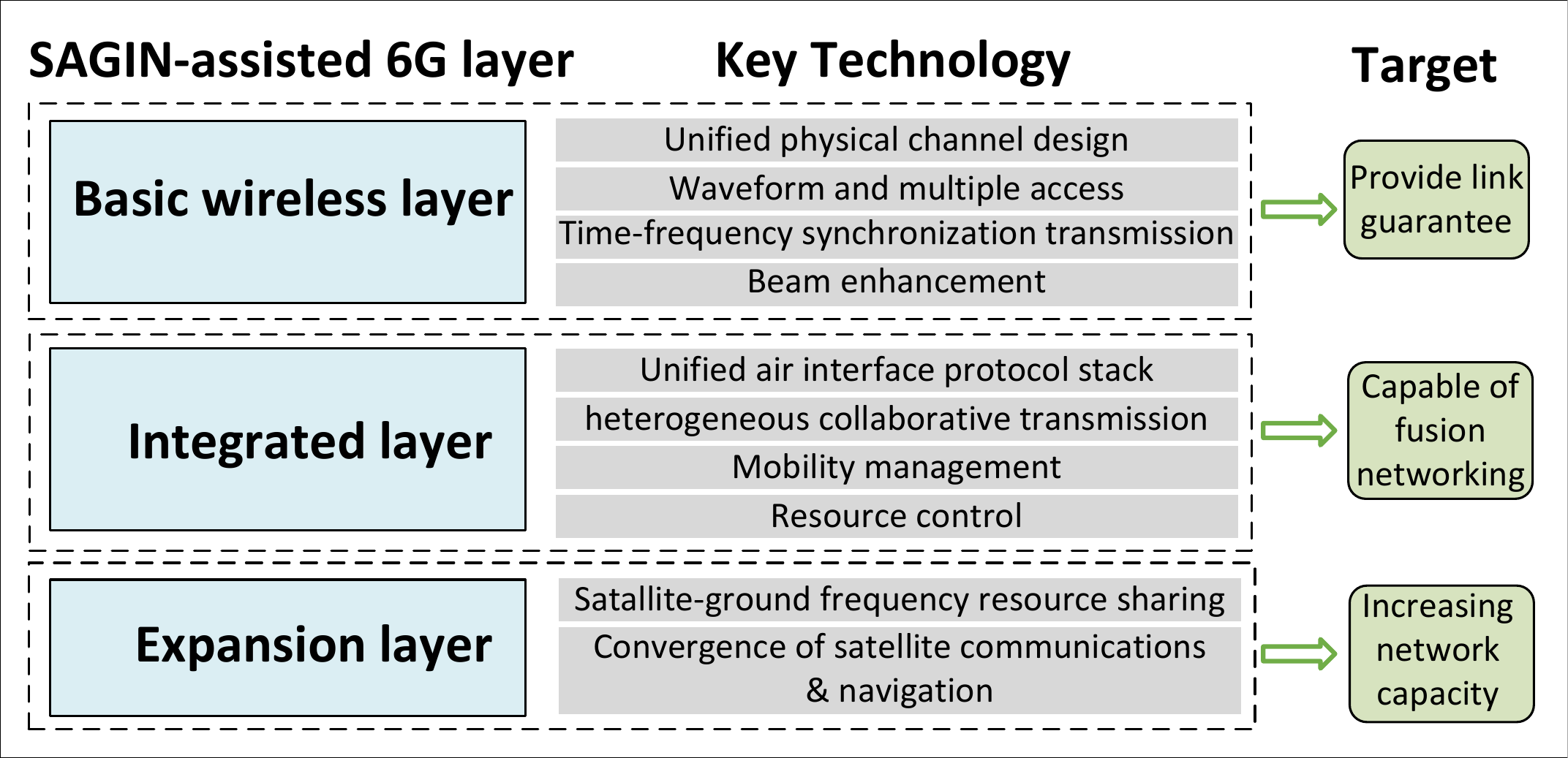}
  \captionsetup{justification=centering}
  \caption{Key technologies of the SAGIN-assisted 6G system.}
  \label{fig:key_tech}
\end{figure*}

To deal with the technical challenges in the design of the SAGIN-assisted 6G system, this paper summarizes a series of key technologies to enable the SAGIN-assisted 6G system to meet the corresponding technical requirements. This SAGIN-assisted 6G can be roughly divided into three layers: basic wireless layer, integrated network layer, and exploration layer, as shown in Fig. \ref{fig:key_tech}. Basic wireless layer mainly includes wireless air interface physical layer transmission technology, laying the technical foundation for space-space integration; integrated network layer focuses on the high level of air interface, network interface and protocol design, requiring SAGIN to meet the joint needs of terrestrial communication and space communication, and provide an integrated design framework; as for the exploration layer, it is a new requirement of the 6G system, which further satisfies frequency fusion and communication navigation fusion on the basis of satellite-ground system integration.

\subsection{Basic Wireless Layer}
The key technologies in the basic wireless layer are shown in Fig. \ref{fig:key_tech}.
\subsubsection{Unified frame structure and physical channel design}
The SAGIN-assisted 6G should adopt a unified frame structure design for various scenarios such as space, air, and ground, and can be parameterized and adaptively formed into time-domain granularities such as wireless frames, subframes, and time slots. By designing flexible and configurable uplink and downlink control channels, various scheduling technologies such as dynamic scheduling, semi-persistent scheduling, and multi-slot scheduling are realized. In view of the characteristics of high delay, Doppler shift, and poor link budget of satellite-ground transmission, the SAGIN-assisted 6G should support time-frequency compensation enhancement technology, HARQ technology and coverage enhancement technology to achieve reliable access and efficient transmission.

\subsubsection{Waveform and multiple access for SAGIN}
To support unified waveform design, the integrated air interface waveform of satellite communication and ground mobile communication will be realized through adaptive configuration based on software parameters. Based on CP-OFDM and DFT-s-OFDM waveforms, the future work is to research the applicability of other waveforms in satellite scenarios.\par
In the future, in addition to supporting orthogonal multiple access technology, non-orthogonal multiple access technology can be considered. In SAGIN, due to the extension of satellite-ground transmission time, the use of non-orthogonal multiple access technology can increase transmission opportunities, increase the number of supported users, and effectively reduce the access delay and data transmission delay of satellite-ground transmission.

\subsubsection{Time-frequency synchronization and reliable transmission}
Faced with the high-speed movement of satellites and terminals, the 5G system has not completely solved the synchronization problem of the non-terrestrial networks air interface. However, the satellite-ground transmission delay is high, the link budget is poor, and the link is easily affected by weather factors. It is also difficult for the SAGIN to support the Time Division Duplexing (TDD) network, and there are inter-cell interference and uplink and downlink timing scheduling problems. It is necessary to research the initial time-frequency synchronization and time-frequency tracking of terminals under high dynamic conditions, and consider various situations such as terminals with positioning capabilities and terminals without positioning capabilities; it is necessary to research low-latency transmission and reliability-enhancing transmission technologies in the SAGIN scenarios and the interference avoidance and timing optimization technology of TDD system in space communication.

\subsubsection{Beam enhancement}
Compared with the beams formed by multiple antennas in terrestrial mobile communications, the beams of satellite communications need to be enhanced. On the one hand, due to the large coverage area of the satellite, it is necessary to consider the trade-off between the number of beams and the power and complexity. On the other hand, the wireless channel of satellite communication is mainly a direct path, and multi-stream transmission is difficult to realize, and usually the payload processing capacity of the satellite is limited, which leads MIMO precoding technology being difficult to implement. Therefore, it is necessary to research the random access mode of satellite communication scenarios and the coordination of various types of beams, the single-user MIMO (SU-MIMO) and multi-user MIMO (MU-MIMO) transmission schemes in satellite communication, and the beam frequency mechanisms for multiplexing and beam interference suppression, etc.













\subsection{Integrated Layer}




The key technologies in Integrated Network Layer are shown in Fig. \ref{fig:key_tech}.\par
In the SAGIN-assisted 6G, all network nodes should define interfaces and protocol stacks in a unified network architecture, support the separation of the user plane and control plane, and support service-oriented and user-centric network architecture design. For the high-speed movement of satellite platforms and air platforms, system can support prediction-based zero-delay and highly reliable handover technology. Based on satellite-ground collaboration and multi-connection technology, satellite-ground collaborative transmission and unified management of resources should be realized. It supports the tailorability and lightweight design of the protocol based on the characteristics of the space environment and the space-borne platform.

\subsection{Exploration Layer}
The key technologies in Exploration Layer are shown in Fig. \ref{fig:key_tech}.\par
\subsubsection{Satellite-ground frequency resource sharing}
With the growth of user service demands, spectrum resources become increasingly scarce. If satellite and terrestrial networks use traditional hard frequency division, the transmission efficiency and frequency utilization will be greatly reduced. For the SAGIN-assisted 6G, frequency resources may no longer be fixedly divided into satellite communications and ground communications, but instead adopt a unified resource allocation strategy. Based on the coordination of terrestrial network station and satellite platforms, interference avoidance and frequency dynamic sharing are realized, and the efficiency of spectrum resource utilization is improved. Spectrum sensing technology needs to be researched to achieve fast interference detection and signal identification, effectively control frequency interference, and dynamically allocate and schedule frequency resources.











\section{Case studies}\label{case}
In this part, we will use one case in Integrated Layer to explain key routing technologies in detail.\par

In SAGIN, congestion and low system performance problems even become more serious compared with terrestrial networks because of the path selection simplification of traditional IP routing protocol. And SAGIN has more equal-cost paths than terrestrial networks. Our work \cite{lan2021exploiting} tries to use path diversity to solve the above problems.\par 
\subsection{Problem Analysis}
Firstly, we use some data to show the severity of the congestion problem in SAGIN.\par 

\subsubsection{Congestion in SAGIN}
In the statistical results \cite{global_internet_map} of TeleGeography, the global Internet capacity in 2020 is about 610 Tbps. We can easily find that most of the world's traffic converges in a few regions like Asia and Europe, especially some big cities such as Singapore, London and New York. Furthermore, the traffic data between the same user and the same server will go through the same path due to the path selection simplification of routing algorithm. When congestion happens, system only can rely on the retransmission mechanism, but this does not solve the problems in the above scenarios.\par 


\subsubsection{Equivalent Paths in SAGIN}
\begin{figure}[htbp]
  \centering
  \includegraphics[width=1.0\linewidth]{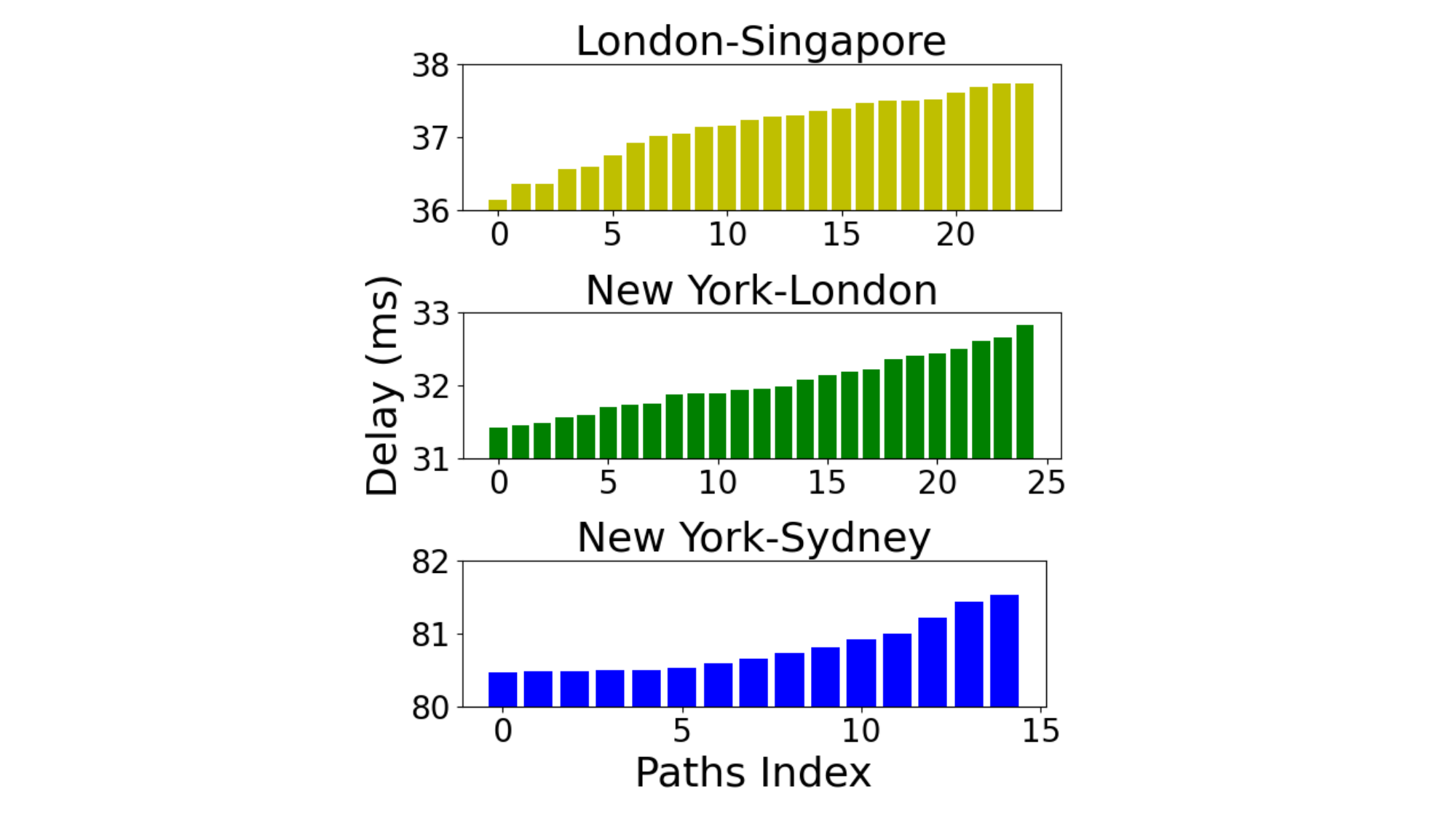}
  \captionsetup{justification=centering}
  \caption{The delay of sub-optimal paths in Starlink.}
  \label{fig:ep}
\end{figure}

Path diversity exists in SAGIN and it is a breakthrough to solve the congestion problem. We have identified a number of sub-optimal paths within SAGIN that have similar delay and bandwidth as the optimal path. These paths are referred to as Equivalent Paths (EPs). In this simulation, three pairs of cities were carefully selected and the delay was meticulously computed across all conceivable paths connecting them. Fig. \ref{fig:ep} shows the shortest dozens of paths in them. It can be observed from Fig. \ref{fig:ep} that there are numerous sub-optimal paths in the satellite network that closely approximate the shortest path.\par

\subsubsection{Difficulties faced by traditional solutions}
In terrestrial networks, there are some famous algorithms proposed for similar problems mentioned above, such as B4 \cite{jain2013b4}. However, such algorithms result in high computational load and long response time due to the high-speed motion of the satellite in SAGIN. Some distributed algorithms, such as ELB \cite{taleb2008explicit}, have improved it. However, since every satellite can only know information from its neighbor, it can only solve sudden network congestion, not congestion for a long period of time.

\subsection{Multi-routing-plane based Flow Scheduling Strategy (MFSS)}
In this work, we designed a system with multi-routing-plane.
Our design features two distinct routing planes, namely the IP routing plane and the auxiliary plane. During periods of low network traffic, all satellites operate on the IP routing plane, thereby keeping the system's computational load at a minimum. However, once the prediction algorithm detects link congestion, the system automatically diverts all subsequent traffic to the auxiliary plane to alleviate the congestion.
Next, the auxiliary plane is activated, in which the MFSS is used to calculate the EPs, and control the traffic to pass the EP to go around the high-load link. Please look for the detail in \cite{lan2021exploiting}.\par

\subsection{Evaluation}
This Section presents a simulation of the MFSS and its ability to enhance system throughput without concomitant increases in delay. Specifically, we conducted a simulation using Starlink, which has 1584 satellites, and employed Systems Tool Kit(STK) software to calculate the propagation delay between satellites. The inter-satellite link bandwidth, denoted as $B_{ISL}$, was set to 400MB. Eighteen cities in Europe, Asia, and North America were selected, and their sending rates were set using a real traffic distribution from the Global Internet Map \cite{global_internet_map}. For comparative purposes, we evaluated the performance of MFSS against the OSPF, B4, and ELB algorithms.\par

\begin{figure}[htbp]
  \centering
  \includegraphics[width=0.8\linewidth]{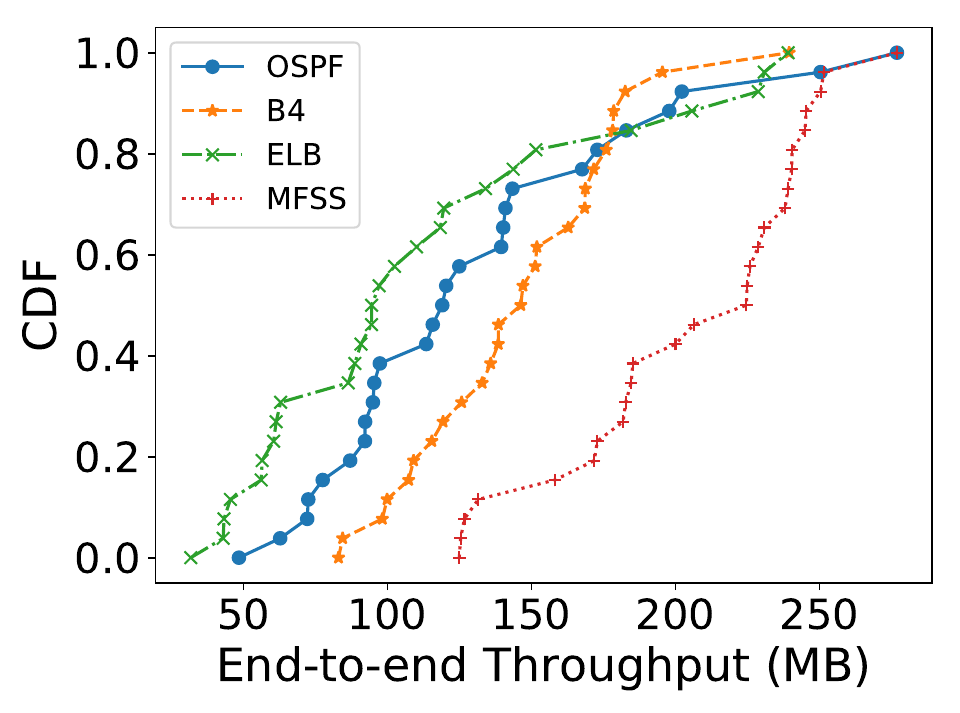}
  \captionsetup{justification=centering}
  \caption{CDF of End-to-End Throughput.}
  \label{fig:result}
\end{figure}

Fig. \ref{fig:result} illustrates the cumulative distribution function (CDF) of all end-to-end throughput measurements collected during the experiment. The results indicate that the adoption of MFSS resulted in a significant increase in the proportion of high-throughput end-to-end connections as compared to other algorithms. For instance, our algorithm achieved over 40\% of the end-to-end connections with a throughput exceeding 200MB, while other algorithms did not exceed 20\%. Additionally, MFSS completely eliminated cases where the throughput was below 125MB, and those cases accounted for up to 70\% when using the ELB algorithm. These findings demonstrate the comprehensive enhancement in system throughput that can be achieved through the use of MFSS in satellite networks.\par

\section{Conclusion}\label{conc}
Driven by multiple needs, 6G should be able to provide all types of users with services in any place, at any time, and in any way. Due to the complementarity between the LEO satellite network and the ground mobile network, building the SAGIN-assisted 6G is the general trend. This paper discussed the motivation, design concept, problem and challenge facing the SAGIN-assisted 6G and proposed a hierarchical architecture and corresponding key supporting technologies.\par 
It is foreseeable that SAGIN will become an important infrastructure in the future to expand the breadth and depth of 6G network services. At the same time, the SAGIN-assisted 6G will provide the world with service capabilities in public safety, emergency communications, social governance, and industrial upgrading.


 


\vspace{11pt}

{
\bibliographystyle{IEEEtran}
\bibliography{IEEEexample}
}

\end{document}